\documentclass[preprintnumbers,amsmath,amssymbm,prd]{revtex4}
\usepackage{epsfig}
\usepackage{graphicx}

\begin{document}
\title{Numerical evidence for universality in the relaxation dynamics of near-extremal Kerr-Newman black holes}
\author{Shahar Hod}
\address{The Ruppin Academic Center, Emeq Hefer 40250, Israel}
\address{ }
\address{The Hadassah Institute, Jerusalem 91010, Israel}
\date{\today}

\begin{abstract}
\ \ \ The coupled gravitational-electromagnetic quasinormal
resonances of charged rotating Kerr-Newman black holes are explored.
In particular, using the recently published numerical data of Dias,
Godazgar, and Santos [Phys. Rev. Lett. 114, 151101 (2015)], we show
that the characteristic relaxation times $\tau\equiv 1/\Im\omega_0$
of near-extremal Kerr-Newman black holes in the regime $Q/r_+\leq
0.9$ are described, to a very good degree of accuracy, by the simple
universal relation $\tau\times T_{\text{BH}}=\pi^{-1}$ (here $Q,
r_+$, and $T_{\text{BH}}$ are respectively the electric charge,
horizon radius, and temperature of the Kerr-Newman black hole, and
$\omega_0$ is the fundamental quasinormal resonance of the perturbed
black-hole spacetime).
\end{abstract}
\bigskip
\maketitle


\section{Introduction}

The influential uniqueness theorems \cite{Isr,Car,Haw1,Rob,Isp} have
revealed that all asymptotically flat stationary black-hole
solutions of the coupled Einstein-Maxwell theory are uniquely
described by the Kerr-Newman spacetime metric
\cite{Chan,Kerr,Newman}. These elegant theorems therefore suggest
that an asymptotically flat perturbed black-hole spacetime would
eventually relax into a stationary Kerr-Newman solution, which is
characterized by only three externally observable conserved
parameters: the black-hole mass $M$, its angular momentum $J\equiv
Ma$, and its electric charge $Q$.

The response of a perturbed black-hole spacetime to external
perturbations is characterized by a unique set of {\it damped}
(complex) oscillations, known as black-hole `quasinormal ringing'.
These damped spacetime resonances are the characteristic `sound' of
the black-hole spacetime itself. They have therefore attracted
considerable interest over the last four decades from both
physicists and mathematicians (see \cite{Nol,Ber,Kon} for excellent
reviews).

The characteristic black-hole quasinormal resonances correspond to
linearized wave fields propagating in the curved spacetime with the
boundary conditions of purely outgoing waves at spatial infinity and
purely ingoing waves crossing the black-hole horizon \cite{Det}.
These physically motivated boundary conditions single out a discrete
spectrum \cite{Noteml} $\{\omega(n;m,l)\}^{n=\infty}_{n=0}$ of
complex (damped) black-hole resonances. The fundamental (least
damped) black-hole quasinormal resonance determines the
characteristic timescale
\begin{equation}\label{Eq1}
\tau_{\text{relax}} \equiv 1/\Im\omega(n=0)\
\end{equation}
for the decay (relaxation) of generic perturbation modes in the
black-hole spacetime.

It should be emphasized that in most situations of physical
interest, the black-hole eigen-frequencies (the characteristic
black-hole quasinormal resonances) are not known in a closed
analytical form . In particular, for most black-hole spacetimes one
is forced to solve the black-hole perturbation equations {\it
numerically} in order to explore the physical properties of the
complex resonance spectra.

Near-extremal Kerr black holes are unique in this respect. In
particular, their fundamental resonances are characterized by the
remarkably simple analytical relation \cite{Hod1,Hod2,Als}:
\begin{equation}\label{Eq2}
\Im\omega(n)=2\pi T_{\text{BH}}(n+{1\over 2}-i\delta)\ \ \ ; \ \ \
n=0,1,2,...\  ,
\end{equation}
where \cite{Noteunit,Notepa}
\begin{equation}\label{Eq3}
T_{\text{BH}}={{r_+-r_-}\over{4\pi(r^2_++a^2)}}
\end{equation}
is the Bekenstein-Hawking temperature of the black hole, and
$\delta$ is the characteristic eigenvalue of the angular black-hole
perturbation equation (the angular Teukolsky equation
\cite{Delta,Teuk}).

The remarkably compact formula (\ref{Eq2}) was derived analytically
in \cite{Hod1,Hod2}. It is worth noting that this formula is {\it
universal} in the sense that it is independent of the spin-parameter
$s$ of the perturbation mode. In particular, the relation
(\ref{Eq2}) provides a quantitative universal description for the
relaxation properties of gravitational ($s=2$), electromagnetic
($s=1$), and scalar ($s=0$) perturbation fields in the rotating {\it
neutral} Kerr black-hole spacetime.

As for the more general case of {\it charged} rotating Kerr-Newman
black holes, the simple relation (\ref{Eq2}) was established
analytically {\it only} for the simplest case of scalar ($s=0$)
perturbation fields. Much less is known about the quasinormal
resonance spectra associated with the coupled
gravitational-electromagnetic perturbations of generic (rotating and
charged) Kerr-Newman black holes. Our limited knowledge about the
quasinormal spectra of generic Kerr-Newman black holes is a direct
consequence of the fact that all attempts to decouple the
gravitational and electromagnetic perturbations of charged rotating
Kerr-Newman black-hole spacetimes have so far failed \cite{Chan}.

Recently, there have been some important numerical studies of the
quasinormal resonance spectra which characterize the charged
rotating Kerr-Newman black-hole spacetimes
\cite{Mar,Pan,Zil,Notearx,Hodarx}. In particular, in a very
interesting work, Dias, Godazgar, and Santos \cite{Dia} have
recently provided detailed numerical results for the characteristic
quasinormal resonances associated with the coupled
gravitational-electromagnetic perturbations of generic Kerr-Newman
black holes.

In the present paper we shall analyze these numerically computed
\cite{Dia} complex black-hole resonances. In particular, we shall
show that the coupled gravitational-electromagnetic perturbations of
near-extremal Kerr-Newman black holes are described extremely well
by the relation (\ref{Eq2}) \cite{Noteeat}. This interesting result
(to be established below) suggests that neutral Kerr black holes and
charged Kerr-Newman black holes share the {\it same} universal
relaxation properties in the near-extremal $T_{\text{BH}}\to0$
regime.

\section{Coupled gravitational-electromagnetic resonances of Kerr-Newman black holes}

Most recently, Ref. \cite{Dia} has provided detailed numerical
results for the fundamental (least damped) quasinormal resonances of
charged rotating Kerr-Newman black holes. We have examined these
numerically computed black-hole resonances in an attempt to reveal a
possible universal pattern which characterizes the relaxation
properties of these charged rotating black-hole spacetimes.

In Table \ref{Table1} we present the black-hole resonances of
near-extremal Kerr-Newman black holes \cite{Dia}. The data shown in
Table \ref{Table1} refer to the fundamental $l=m=2$ \cite{Noteml}
gravitational-electromagnetic black-hole quasinormal resonances with
$M\Im\omega_0=0.01$. In particular, we display the dimensionless
ratio $\Im\omega_0/\pi T_{\text{BH}}$ for various values of the
dimensionless black-hole charge parameter $Q/r_+$.

Interestingly, the numerical data presented in Table \ref{Table1}
provide compelling evidence that the relaxation rates of generic
(that is, charged and rotating) Kerr-Newman black holes in the
regime $Q/r_+\leq0.9$ \cite{Notether} are governed by the simple
universal relation
\begin{equation}\label{Eq4}
\Im\omega(n=0)\to\pi T_{\text{BH}}\ \ \ \text{as}\ \ \
T_{\text{BH}}\to 0\ .
\end{equation}

\begin{table}[htbp]
\centering
\begin{tabular}{|c|c|}
\hline $\ \ Q/r_+\ $ & \ \ ${{\Im\omega_0}/{\pi T_{\text{BH}}}}\ \  $\ \ \ \\
\hline
\ \ 0.2\ \ \ &\ \ 0.978\\
\ \ 0.3\ \ \ &\ \ 0.987\\
\ \ 0.4\ \ \ &\ \ 0.972\\
\ \ 0.5\ \ \ &\ \ 0.980\\
\ \ 0.6\ \ \ &\ \ 0.945\\
\ \ 0.7\ \ \ &\ \ 0.987\\
\ \ 0.8\ \ \ &\ \ 0.972\\
\ \ 0.9\ \ \ &\ \ 0.999\\
\hline
\end{tabular}
\caption{Quasinormal resonances of near-extremal charged rotating
Kerr-Newman black holes. The data shown refer to the fundamental
$l=m=2$ gravitational-electromagnetic perturbation mode with
$\Im\omega_0=0.01M^{-1}$ \cite{Dia}. We display the dimensionless
ratio $\Im\omega_0/\pi T_{\text{BH}}$ for various values of the
dimensionless black-hole charge parameter $Q/r_+$, where
$T_{\text{BH}}={{(r_+-r_-)}/{4\pi(r^2_++a^2)}}$ is the temperature
of the corresponding Kerr-Newman black hole. One finds that, for
near-extremal Kerr-Newman black holes in the regime $Q/r_+\leq 0.9$
\cite{Notether}, the quasinormal resonance spectra are
characterized, to a good degree of accuracy, by the universal
relation $\Im\omega_0/\pi T_{\text{BH}}\to1^{-}$.}
\label{Table1}
\end{table}

\section{Summary}

Our scientific knowledge about the resonance spectra of generic
(that is, charged and rotating) Kerr-Newman black holes is not as
good as our knowledge about the corresponding quasinormal spectra of
rotating neutral Kerr black holes. This unsatisfactory state of
affairs stems from the fact that all attempts to decouple the
gravitational and electromagnetic perturbations of the Kerr-Newman
black-hole spacetime have so far failed \cite{Chan}.

Thus, in order to study the coupled gravitational-electromagnetic
quasinormal spectra of generic Kerr-Newman black holes, one is
forced to solve numerically a set of two coupled partial
differential equations \cite{Chan}. In this paper we have analyzed
these numerically computed \cite{Dia} black-hole resonances. In
particular, we have provided compelling evidence that the
characteristic relaxation times, $\tau\equiv 1/\Im\omega_0$, of
perturbed Kerr-Newman black-hole spacetimes in the regime $Q/r_+\leq
0.9$ \cite{Notether} are characterized by the compact universal
relation
\begin{equation}\label{Eq5}
\tau\times T_{\text{BH}}\to\pi^{-1} \ \ \ \text{as}\ \ \
T_{\text{BH}}\to 0\  ,
\end{equation}
where $T_{\text{BH}}$ is the Bekenstein-Hawking temperature of the
Kerr-Newman black hole [see Eq. (\ref{Eq3})]. The relation
(\ref{Eq5}) suggests that neutral Kerr black holes and charged
Kerr-Newman black holes share the {\it same} universal relaxation
properties in the near-extremal $T_{\text{BH}}\to0$ regime.

We believe that it would be physically interesting (and
mathematically challenging) to find an analytical explanation
\cite{Noteexp} for the numerically suggested universal behavior
(\ref{Eq5}) which characterizes the near-extremal Kerr-Newman
black-hole spacetimes.

\bigskip
\noindent
{\bf ACKNOWLEDGMENTS}

This research is supported by the Carmel Science Foundation. I would
like to thank Yael Oren, Arbel M. Ongo, Ayelet B. Lata, and Alona B.
Tea for helpful discussions.


\bigskip

\begin{thebibliography}{99}

\bibitem{Isr} W. Israel, Phys. Rev. {\bf 164}, 1776 (1967); Commun. Math. Phys. {\bf 8},
245 (1968).

\bibitem{Car} B. Carter, Phys. Rev. Lett. {\bf 26}, 331 (1971).

\bibitem{Haw1} S. W. Hawking, Commun. Math. Phys. {\bf 25}, 152 (1972).

\bibitem{Rob} D. C. Robinson, Phys. Rev. D {\bf 10}, 458 (1974); Phys. Rev. Lett. {\bf 34},
905 (1975).

\bibitem{Isp} J. Isper, Phys. Rev. Lett. {\bf 27}, 529 (1971).

\bibitem{Chan} S. Chandrasekhar, {\it The Mathematical Theory of Black
Holes}, (Oxford University Press, New York, 1983).

\bibitem{Kerr} R. P. Kerr, Phys. Rev. Lett. {\bf 11}, 237 (1963).

\bibitem{Newman} E. T. Newman, R. Couch, K. Chinnapared, A. Exton,
A. Prakash, et. al., J. Math. Phys. {\bf 6}, 918 (1965).

\bibitem{Nol} H. P. Nollert, Class. Quantum Grav. {\bf 16}, R159 (1999).

\bibitem{Ber} E. Berti, V. Cardoso and A. O. Starinets, Class. Quant. Grav. {\bf 26}, 163001 (2009).

\bibitem{Kon} R. A. Konoplya and A. Zhidenko, Rev. Mod. Phys. {\bf 83}, 793 (2011).

\bibitem{Det} S. L. Detweiler, in Sources of Gravitational Radiation, edited by
L. Smarr (Cambridge University Press, Cambridge, England, 1979).

\bibitem{Noteml} Here $m$ and $l$ are respectively the azimuthal harmonic index and
the spheroidal harmonic index of the black-hole perturbation mode.

\bibitem{Hod1} S. Hod, Phys. Rev.D {\bf 78}, 084035 (2008) [arXiv:0811.3806].

\bibitem{Hod2} S. Hod, Phys. Lett. B {\bf 715}, 348 (2012)
[arXiv:1207.5282].

\bibitem{Als} See also: S. Hod, Phys. Rev. D {\bf 75}, 064013 (2007)
[arXiv:gr-qc/0611004]; S. Hod, Class. and Quant. Grav. {\bf 24},
4235 (2007) [arXiv:0705.2306]; A. Gruzinov, arXiv:gr-qc/0705.1725;
S. Hod, Phys. Rev. D {\bf 78}, 084035 (2008) [arXiv:0811.3806]; A.
Pesci, Class. Quantum Grav. {\bf 24}, 6219 (2007); S. Hod, Phys.
Lett. B {\bf 666}, 483 (2008) [arXiv:0810.5419]; S. Hod, Phys. Rev.
D {\bf 80}, 064004 (2009) [arXiv:0909.0314]; S. Hod, Phys. Lett. A
{\bf 374}, 2901 (2010) [arXiv:1006.4439]; S. Hod, Phys. Rev. D. {\bf
84}, 044046 (2011) [arXiv:1109.4080]; S. Hod, Phys. Lett. B {\bf
710}, 349 (2012) [arXiv:1205.5087]; S. Hod, Phys. Lett. B {\bf 715},
348 (2012) [arXiv:1207.5282]; S. Hod, Phys. Rev. D {\bf 88}, 084018
(2013) [arXiv:1311.3007]; S. Hod, Phys. Lett. B {\bf 747}, 339
(2015) [arXiv:1507.01943].

\bibitem{Noteunit} We use natural units in which $G=c=\hbar=1$.

\bibitem{Notepa} Here $M, Ma, Q$,
and $r_{\pm}=M\pm(M^2-a^2-Q^2)^{1/2}$ are respectively the mass,
angular momentum, electric charge, and horizon-radii of the black
hole.

\bibitem{Delta} The parameter $\delta$ is closely related to the
angular-eigenvalue $\lambda$ of the angular Teukolsky equation, see
\cite{Teuk} for details [see, in particular, equations (2.7) and
(6.3) of \cite{Teuk}].

\bibitem{Teuk} S. A. Teukolsky and W. H. Press, Astrophys. J. {\bf 193}, 443
(1974).

\bibitem{Mar} Z. Mark, H. Yang, A. Zimmerman, and Y. Chen, Phys. Rev. D {\bf 91}, 044025 (2015).

\bibitem{Pan} P. Pani, E. Berti, and L. Gualtieri, Phys. Rev. Lett. {\bf
110}, 241103 (2013); P. Pani, E. Berti, and L. Gualtieri, Phys. Rev.
D {\bf 88}, 064048 (2013).

\bibitem{Zil} M. Zilh\~ao, V. Cardoso, C. Herdeiro, L. Lehner, and U. Sperhake,
Phys. Rev. D {\bf 90}, 124088 (2014).

\bibitem{Notearx} Using the numerical results of \cite{Zil}, it was shown in \cite{Hodarx} that
charged Kerr-Newman black holes (like neutral Kerr black holes
\cite{Hod1}) are characterized by the simple relation $\Re\omega\to
m\Omega_{\text{H}}$ in the near-extremal $M^2-Q^2-a^2\to 0$ limit,
where $\Omega_{\text{H}}=a/(r^2_++a^2)$ is the angular-velocity of
the black-hole horizon.

\bibitem{Hodarx} S. Hod, The Euro. Phys. Jour. C (Letter) {\bf 75}, 272 (2015)
[arXiv:1410.2252].

\bibitem{Dia} O. J. C. Dias, M. Godazgar, and J. E. Santos, Phys. Rev. Lett. {\bf 114}, 151101 (2015).

\bibitem{Noteeat} It is worth emphasizing again that, for charged rotating Kerr-Newman black holes,
the simple relaxation spectrum (\ref{Eq2}) was derived analytically
only for the particular case of scalar ($s=0$) perturbation modes.
This restriction to the scalar case is a direct consequence of the
fact that all attempts to decouple the gravitational and
electromagnetic perturbations of generic (that is, charged and
rotating) Kerr-Newman black holes have so far failed \cite{Chan}.

\bibitem{Notether} This is the regime of black-hole electric charges studied numerically in
\cite{Dia}.

\bibitem{Noteexp} It is worth emphasizing that Ref. \cite{Hod1} has already provided
an analytical explanation for the simple asymptotic behavior
(\ref{Eq5}) in the specific case of {\it neutral} near-extremal Kerr
black holes.



\end{thebibliography}
\end{document}